\documentclass[12pt,preprint]{aastex}
 \let\b=\beta   \let\e=\epsilon
  \let\q=\theta  
    
\let\s=\sigma    \let\c=\chi

 \let\W=\Omega

\def\grays{$\gamma$--rays}
\usepackage{graphics}    
\newcommand{\be}{\begin{equation}}
\newcommand{\ee}{\end{equation}}

\newcommand{
\chandra}{{\it Chandra} }

\shorttitle{Compact nebulae of Vela and PSR B1706-44}
\shortauthors{Sefako R. R. \& de Jager, O. C}

\def\laeq{\;\raise.2ex\hbox{$<$}\kern-.75em\lower.9ex\hbox{$\sim$}\;}
\def\gaeq{\;\raise.2ex\hbox{$>$}\kern-.75em\lower.9ex\hbox{$\sim$}\;}

\begin{document}

\title{Constraints on pulsar magnetospheric and wind parameters for the 
compact nebulae of Vela and PSR B1706-44.}

\author{Ramotholo R. Sefako and Ocker C. de Jager}
\affil{Unit for Space Physics, Potchefstroom University, 
Potchefstroom 2520, South Africa;\\ Email: okkie@fskocdj.puk.ac.za}
\begin{abstract}
We present a new model for the synchrotron compact nebular 
emissions for Vela and
PSR B1706-44, and derive fundamental pulsar/plerion parameters such as the pair 
production multiplicity, $M$, and wind magnetization parameter, $\sigma$.
The pair cascade above the pulsar polar cap, combined with the
energy from the pulsar wind, injects a pair plasma into the 
surrounding environment. This wind, consisting of particles
and fields, is shocked by the environment, resulting in synchrotron
emission from ``thermalized" pairs in a compact nebula. The broad band nebular
spectrum depends crucially on the spindown power, distance,
pair multiplicity, pulsar wind shock radius 
and $\sigma$ at this shock. We construct such a model  
for the particle spectra in the compact X-ray nebulae of Vela and
PSR B1706-44.  Fits to the multiwavelength spectra of these
sources indicate that $300<M<1000$, whereas $0.05<\sigma<0.5$ for Vela. 
The same $\sigma$ interval (as for Vela) was independently 
derived from the radial gradients of the X-ray compact nebular emission
from both Vela and PSR B1706-44, giving us confidence in our results.
The $M$ we derive
for Vela is too small to be explained by curvature losses in the open magnetosphere
of Vela, but the presence of optical pulsed photons could modify the
predicted multiplicity. 
\end{abstract}   
\keywords{pulsars: individual (PSR B1706-44, Vela) --- radiation mechanisms: non-thermal ---  X-rays: stars }


\section{Introduction}
\label{pler_intro}
Plerions are pulsar--driven nebulae showing broadband synchrotron
emission which is usually steady as a result of the extended size
of these sources.
The word ``plerion'' (Greek for ``wind bag'')
indicates that the nebula is powered by a central source,
mostly a young to middle aged pulsar.
The study of plerions is important since it allows us to probe the particle
output from the pulsar magnetosphere into the interstellar medium. 

Basic pulsar physics teaches that the Goldreich-Julian current (Goldreich \& Julian 1969)
is multiplied through an electromagnetic cascade (Sturrock 1971), with curvature
radiation decreasing the mean energy per particle along the cascade process. 
General relativistic frame dragging was found to be important in the creation
of the accelerating potential (Muslimov \& Tsygan, 1990, 1992;
Muslimov \& Harding 1997), while inverse
Compton scattering (ICS) was found to be important (or even the dominant process)
for high energy photon production (Sturner, Dermer \& Michel 1995;
Luo 1996; Zhang et al. 1997). In fact, Zhang \& Harding (2000) found that the 
inclusion of the ICS process
could explain the fundamental features of the observed $\gamma$-ray spectra
from energetic pulsars. Finally, Hibschman \& Arons (2001) 
combined the Muslimov \& Tsygan (1990, 1992) accelerating potential with curvature
and ICS processes to give analytical expressions for the pair production
(or pair multiplicity, $M$) rates. 

The same number of pairs that escape through the
pulsar light cylinder should eventually enter the plerion and be visible
as non-thermal radiation after these pairs were ``thermalized". The only problem,
however, is that the final pair energy emergent from the light cylinder
may be too low to explain the non-thermal high energy emission from the 
large (parsec-scale) emission volumes associated with plerions.
Thus, whereas the total particle rate to be injected into the plerion
is expected to remain conserved, particle acceleration should take place
beyond the light cylinder, so that the high photon energies and their associated
intensities may be explained. Rees \& Gunn (1974) suggested that energy should be
converted from the electromagnetic component in the pulsar wind to the pair plasma.
Kennel \& Coroniti (1984a) refined this model by deriving 1-D 
magnetohydrodynamic (MHD) flow solutions, 
which depends on the degree of such energy transfer, characterized by the
``magnetization-" or ``$\sigma$-parameter", defined to be the ratio of
electromagnetic energy flux relative to the particle energy flux. 
A low value of $\sigma$ would imply a strong pulsar-wind shock, whereas
a value near unity would imply a weak shock. A value $\sigma \gg 1$ would imply
that a discontinuity (or jump in the wind parameters) or shock should be
absent.
  
Whereas attempts have been made to model the post shock MHD flow in
plerions (Kennel \& Coroniti 1984a,b), the exact process of conversion of 
electromagnetic energy to particle energy (i.e. a reduction in $\sigma$)
between the light cylinder and the pulsar wind shock appears to 
remain unanswered (Begelman 1998;
Lyubarsky \& Kirk 2001), although a solution for such an increase in particle
energy beyond the pulsar light cylinder has been forwarded by
Contopoulos \& Kazanas (2002).
   
MHD studies of the Crab Nebula by Kennel \& Coroniti (1984a)
indicated that the ratio
$\sigma$ near the pulsar wind shock radius, $r_{\rm s}$ (i.e. the
radius where the pulsar ram pressure balances the external
ambient pressure), may be as low as 0.003. Recently, de Jager (2001)
has shown that the TeV data from the Crab Nebula
are consistent with this prediction, even though the experimental
uncertainties are too large to confirm
$\sigma\sim 0.003$ conclusively for the Crab Nebula.    
Furthermore, the direct detection of unpulsed TeV {\grays} from a few plerions
such as the Crab (Weekes et al. 1989), PSR B1706--44 (Kifune et al. 1995),
Vela (Yoshikoshi et al. 1997) and PSR B1509--58 (Sako et al. 2000)
also proves that electrons with energies well
above a few TeV must be present in such sources.

A fresh approach is required to constrain $\sigma$ empirically for pulsar winds,
but it will be found that this constraint also depends on the pair
creation multiplicity, $M$, in the pulsar magnetosphere, if an empirical
approach is used. Rather than attempting to determine $\sigma$
in the radiation zone from fundamental principles (which is difficult
to achieve), we will constrain $M$ and $\sigma$ simultaneously
from multi-wavelength data. This is achieved by first performing
a study of the broadband spectrum in the radiation zone, from
which empirical constraints can be derived
for the parameters in question using a robust model.
The basic equations, which determine the particle and hence radiation
spectra, will be presented in Section \ref{pler_modl}.  

The most important consideration when attempting to constrain fundamental parameters
such as $M$ and $\sigma$ is that we have to select a
detectable emission volume for which the number of model free parameters is a
minimum. In this case, we will
restrict ourselves to observations taken from
the region behind the pulsar wind shock (the radiation zone as observed in X-rays
and hopefully at other wavelengths as well),
where the convection speed (with $v\sim c$) is expected
to dominate diffusion effects (diffusion models usually have more free
parameters). In this region the parametric equations for the
post shock magnetic field strength and convection speed
were derived from the 1-D MHD approach of Kennel \& Coroniti (1984a), with 
$\sigma$ the only explicit free parameter, apart from the spindown power, 
distance and pulsar
wind shock radius (to be determined from observations), 
whereas the radius from the pulsar is the variable in this 1-D model.

The employment of a 1-D MHD flow solution is not strictly speaking
correct, since pulsar winds clearly show at least a 2-D geometry, with
equatorial and polar zones (Aschenbach \& Brinkmann 1975; Hester et al. 1995;
Helfand et al. 2001). In fact, Bogovalov (2001) commented that the $\sigma$
parameter corresponding to the equatorial zone may be smaller than the
corresponding parameter in the polar zone, which would explain the bright
X-ray torus observed from the Crab Nebula. The validity of using a 1-D flow
solution for the toroidal emission region will be revisited in subsequent Sections.

High resolution imaging and spectroscopy of the 
compact nebulae of Vela (Helfand et al. 2001) and PSR B1706-44
(Finley et al. 1998; Gotthelf, Halpern \& Dodson 2002; 
Dodson \& Golap 2002) are available for X-rays, 
whereas only \"Ogelman, Koch--Miramond
and Auri\'ere (1989) attempted to image the Vela compact nebula
in optical. The latter authors have shown that the optical 
Vela compact nebula is comparable in size to the X-ray
compact nebula and the X-ray power law spectrum 
extends into
the optical domain. De Jager, Harding \& Strickman (1996a) 
have also shown that the same hard spectrum
seems to be extending to 0.4 MeV, resulting in an unbroken power
law spectrum over nearly six decades in energy. \"Ogelman et al. (1989) 
also showed that the optical and X-ray compact nebulae are
comparable in size, although no detailed map was produced.
It will be shown that such a spectrum (determined from X--ray observations) must
be constrained by radio and/or optical observations, otherwise it would be 
difficult to constrain $\sigma$ and $M$ simultaneously. 
Whereas $\sigma$ determines the total energy in the radiating
component, $M$ determines the spectral bandwidth -- 
a large $M$ distributes
the energy to many particles, resulting in a reduction in the
mean energy per electron. The consequence of this is that the compact 
nebula may be
seen down to optical frequencies, whereas a smaller $M$ confines the
power to less particles, which may result in a spectral turnover
between optical and X-rays.

Interpolating between typical radio and X-ray
plerionic spectra (for a fixed volume) indicates that the total optical intensity
should be relatively large compared to the optical brightness
of the pulsar itself (e.g. Urama \& Okeke 1998),
but the large size of the nebula (compared to a typical CCD pixel size) makes the signal
to noise ratio per pixel small, given the background detector noise and
starlight contribution. Despite this empirical constraint, it
is still important to derive empirical limits for the optical
emission associated with the compact X-ray nebula and possible
extended nebula (larger than the compact nebula), which are
required to explain the unpulsed TeV emission from a source
such as PSR B1706-44 (Aharonian, Atoyan \& Kifune, 1997). 
  
The available multiwavelength data of the compact plerions of Vela and
PSR B1706-44 will be combined for further studies in Sections \ref{pler_vela} 
and \ref{pler_1706}, respectively. Section \ref{pler_modl}  presents a robust model for 
the synchrotron emissivity
of the equatorial zone associated with the plerions of Vela and PSR B1706-44.
Confidence contours for the parameters $M$ and $\sigma$ are derived in Section
\ref{pler_pair}, given the multiwavelength spectra, pulsar parameters, as well as
geometrical constraints as inputs. Finally, Section \ref{pler_disc} will discuss the main 
points of this paper.

\section{The Vela compact nebula}
\label{pler_vela}

Vela pulsar is a middle aged 
pulsar (characteristic age, 11.5 kyr) with a relatively large spindown power 
($\dot{E}=6.9\times10^{36}\; {\rm erg~s^{-1}}$) and rotation period of 89~ms. 
This pulsar has been observed to show pulsed emission in 
radio, optical, X--rays and GeV energies (Thompson et al. 1996), and
the super--exponential cutoff below 10 GeV (Nel \& de Jager 1995)
probably explains why 
no pulsed TeV emission was detected. It was one of the first 
isolated neutron stars to be observed optically and it is known to
glitch quite often. Mignani \& Caraveo (2001), using HST data, have shown that the pulsar 
spectral distribution is dominated by a flat power-law continuum with no spectral
turnover to wavelengths exceeding 7000 \AA.  
Recent studies of the Vela remnant by  Cha,
Sembach \& Danks (1999) put Vela pulsar at just 250~pc away, whereas the 
Hubble Space Telescope (HST) astrometric parallax measurement 
by Caraveo et al.
(2001) gives the latest distance measurement to Vela as $295\pm50$~pc. 
The distance measurement by Caraveo et al. (2001) will be adopted
in this study.   

Helfand et al. (2001) show the \chandra HRC X--ray image centered on the 
Vela pulsar position, indicating a point source located at the pulsar position, 
and a jet
and counter--jet southeast and northwest of the pulsar point,
respectively, along the direction of the jet. These jet 
structures appear to be aligned with the direction of the
proper motion of the pulsar roughly towards the northwest (Caraveo et al.
2001).  
A nebular arc structure, consisting of bright and faint
regions, perpendicular to the jet direction 
is also visible. 
By balancing the ambient gas pressure and pulsar wind pressure, Helfand et al.
(2001) calculated a radius of $r_s=52.''7$ for the pulsar wind shock. However, 
inner MHD shocks are also present, which appear to be confined to a 
region of about $0.'5$ radius.
 
\"Ogelman et al. (1989) detected an 
optical counterpart which is similar in size to the X-ray compact nebula. 
Furthermore, the optical flux appears to be  
consistent with the extrapolated X-ray compact nebular spectrum.
Lewis et al. (2001) imaged the radio counterpart of this compact nebula, showing
two radio lobes which are probably perpendicular to the pulsar spin axis.
However, the radio lobes start at the edge of the X-ray emission, extending to
a size three times the radius of the compact X-ray nebula. It is, thus, clear
that the particles responsible for the radio and X-ray emissions do not
belong to the same spectral/spatial population, although a single
transport equation should be able to describe the relation between the X-ray
and radio nebulae.

The physics relevant to the compact nebula have been discussed
by a number of authors in the past (\"{O}gelman et al. 1989;
de Jager et al. 1996a,b; Pavlov et al. 2001; 
Helfand et al. 2001), with Helfand et al. (2001) suggesting an equatorial
Crab-like origin for the bright X-ray arcs. However, Radhakrishnan \& Deshpande (2001) 
revisited the interpretation of the radio polarization in Vela, resulting in
a natural explanation of the two bright X-ray arcs as arising from a double
cone with half angle $\sim 70^{\circ}$ (the angle between the magnetic and
spin axis), with the two cones originating from the
two magnetic poles. The two bright arcs and subsequent radio lobes, therefore,
originate from the sweeping of the two magnetic poles (nearly orthogonal
rotator) around the spin axis. Several questions still remain unanswered with this
model. 

The X-ray lobes appear to be distorted towards the pulsar `birthplace' (obtained from 
pulsar proper motion and its canonical age), which may indicate that particles are 
transported
in a wave towards this position, where excess TeV $\gamma$-ray emission was
detected by Yoshikoshi et al. (1997). More sensitive TeV observations are
however required to confirm this finding. Whereas an excess of X-ray
emission was detected from the birthplace (Harding, de Jager \& Gotthelf 1997), no 
evidence of any such excess in optical has been observed. 

Despite the different views on the origin of the X-ray arcs in the Vela compact nebula,
we will construct a robust model for particle emission in the compact nebula
which can account for both viewpoints. The only concern for this paper is to
have a wide spectral coverage which can constrain model parameters. Such 
parameters are useful for any assumed model (polar or equatorial) for the 
bright X-ray arcs.       

The radio limit for Vela
implies a turnover in the synchrotron spectrum of the compact
nebula. This is expected, otherwise the number of electrons
required to have the synchrotron spectrum extending all the
way to radio frequencies, would constrain the pulsar/nebular 
parameters severely. The constraints on the compact nebular
spectrum will be exploited in Section \ref{pler_pair} when the pulsar/nebular
parameters are constrained.

\section{The compact nebula of PSR B1706--44}
\label{pler_1706}

PSR B1706--44 is a middle-age Vela--like pulsar (spin--down age 
17.5~kyr), with a period of 102 ms and a spin--down power of
\.{E}~=~3.4~x~10$^{36}$~erg/s. It was discovered during a 20~cm 
radio pulsar 
survey of the southern Galactic plane (Johnston et al. 1992), and later
detected in soft X--rays during the ROSAT mission (Becker, Brazier \& Tr\"{u}mper
1995; Becker et al. 1992). 
A dispersion based distance measure of Taylor \& Cordes (1993)
places PSR B1706--44 at approximately 1.8~kpc, whereas 21 cm hydrogen line
absorption data taken by Koribalski et al. (1995) imply a distance range
of 2.4 to 3.2 kpc. 

Finley et al. (1998) discovered a Vela-like compact X-ray nebula associated with
PSR B1706-44, whereas Frail et al. (1994) discovered 
a radio synchrotron nebula associated
with the pulsar, with a radius of $\sim 2$~arcmin at 20~cm wavelength. 
A similar radio synchrotron 
detection was recently presented by Giacani et al. (2001), 
confirming a radio
nebular size of about $3.'5\times2.'5$ around the pulsar position. Thus, the
general similarity with Vela has been established: the radio compact nebula 
of PSR B1706-44 is also larger compared to its X-ray compact nebula. If the systems
are similar, the size of the PSR B1706-44 compact nebula should be a factor ten
smaller, given the ten times larger distance. Scaling the radii of the 
bright X-ray arcs (22 to 29 arcsec) of Vela to the distance of PSR B1706-44, gives radii
between 2 and 3 arcsec (assuming the same confining pressure and pulsar 
spindown power). In fact Gotthelf et al. (2002) and Dodson \& Golap (2002)
derived a Gaussian of $\sim 1.1$ arcsec, with a tail extending about 5 arcsec
towards the north westerly direction - towards the center of SNR G343.1-2.3,
which makes the association with this SNR more likely (Dodson \& Golap 2002).
Given the revised morphology of the SNR, it is then possible
that the transverse speed may be close to the value of 100 km/s as measured
by Johnston, Nicastro \& Koribalski (1998).
    
This pulsar was also identified as a pulsed GeV source by the EGRET 
instrument (Thompson et al. 1992). Furthermore,
very high energy $\gamma$--ray observations above 1~TeV from CANGAROO 
(Kifune et al. 1995; Kifune 1997) and above 0.3~TeV from the Durham 
(Chadwick et al. 1997) VHE experiments have
confirmed the existence of unpulsed radiation from this source. 
The CANGAROO detection is a source with angular extend
not exceeding $\sim 0.^{\circ}2$, whereas the angular extend from 
Durham observations implies a radius not exceeding $\sim 0.^{\circ}15$
(P. Chadwick 2001, personal communication). Even though the compact nebula
is a few arcseconds in size, the TeV nebula has to be much larger in size
(de Jager 1995; Aharonian et al. 1997; Harding \& de Jager 1997). The 
explanation for this is beyond the scope of this paper, given its objectives,
but the pulsar/pulsar wind parameters derived from this paper will be tested
against robust models and observations of TeV $\gamma$-rays.
  
Chakrabarty \& Kaspi (1998)  
gave a red--band limit of $R\gaeq 18$ to the pulsar. A 3-sigma  
limit to the pulsar magnitude of $V=24.5$ was given by Lundqvist et al. (1999)
and this is consistent with the theoretical predictions of $V=24.12$ (e.g.
Urama \& Okeke 1998, and references therein).  Mignani et al. (1999) got a 
limit of $V\gaeq 27.5$ using similar data to Lundqvist et al. (1999). 
It is, therefore, clear that the PSR B1706--44 pulsar is extremely faint compared
to the Vela pulsar. No previous attempts have been recorded to
search for a compact nebula in optical, and our attempts to search for this were
also unsuccessful (Sefako 2002). 

Given the RMS X-ray size of 1.1 arcsec,
the trail extending not more than 5 arcsec, and the improved radio timing position
of the pulsar (Wang et al. 2000), we may have to revisit the observations of 
Mignani et al.
(1999) and Lundqvist et al. (1999) to rederive a V-band limit for the optical
compact nebula corresponding to the X-ray image.

With the improved pulsar position of Wang et al. (2000), we find that the X-ray 
trail 
is pointing away from the contaminating effect of star 1 listed by Mignani et al.
(1999), but the tail of the PSF of star 1 touches the X-ray contours corresponding
to the bow shock at the southern part of the compact nebula (see Figure 3
of Dodson \& Golap, 2002). Thus, the plerion, if visible, should appear as a 
$1''$ -- $2''$ radius
object near the northern edge of the PSF of star 1. Whereas the bow shock may
be contaminated, we should at least see the rest of the plerion well separated
from the PSF of star 1. No such structure is seen, and an object with $V\sim 24.5$ mag
and radius $1''$ -- $2''$ should be visible on the plate. We will, therefore, 
employ an upper
limit of $V>24.5$ mag for the optical counterpart of the compact X-ray 
nebula. The unabsorbed magnitude should be $V> 21.2$ assuming an extinction of
$A_v=3$. This limit is still not constraining given the hard X-ray spectrum of
the compact nebula of the source.

\section{A 2D model for compact nebular emission}
\label{pler_modl}
Both the Crab and Vela compact nebular emissions appear to
be toroidal. \chandra observations of 
PSR B1706--44 also show resolved emission, although the
evidence for a torus is not clear. Although Vela and PSR B1706-44
are similar in age and spindown power, the ten times
larger distance for PSR B1706-44 compared to Vela makes
the unambiguous identification of the appropriate geometry difficult.
We will, however, assume that most of the X-ray emission
from PSR B1706-44 is also in the form of toroidal emission.
However, if the geometry is similar to that proposed by Radhakrishnan \& 
Deshpande (2001, hereafter RD01) for Vela,
the numerical results derived should hold, but interpreted in terms
of the RD01 model. 

This section will concentrate on the compact nebular
emission resulting from the convection of electrons in the
pulsar wind, with most of the emission arising from the interaction
of the pulsar wind shock with the external gas pressure -- whether this
is static due to the thermal gas pressure, or dynamic due to proper motion. 
This interaction
results in the randomization of pitch angles, resulting in the 
observations of bright arcs -- at least for Vela (Helfand et al. 2001). In the
case of Vela, the thermal gas pressure correctly predicts the
size of the compact nebula. Helfand et al. (2001) estimated a size of 
$r_s\sim 33''$ given the thermal pressure 
($P_{gas}=8.5\times10^{-10}\; {\rm erg\; cm^{-3}}$) derived from the
observation of thermal X--rays in the inner regions of
the Vela SNR. 
In the case of the compact nebula of PSR B1706--44, 
Dodson \& Golap (2002) determined the particle density of proper motion which
predicts the arcsecond size of the compact nebula. 
Vela and PSR B1706-44 will, otherwise, be treated similarly in the
model presented below.

\subsection{A robust model for particle injection}
A general model is assumed for the injection spectrum 
in the radiation zone $Q(E)$, evaluated at the pulsar wind shock
where most of the compact nebular emission is seen. This spectrum
can reproduce the spectral turnover below an energy $E_o$
to account for the fact that the compact nebular X--ray spectra
cannot extrapolate to radio frequencies in the case of Vela. 
The fastness of the turnover is determined by the
parameter $f$, which will be left free, although this
value is weakly constrained by observations near the
turnover. The spectral index below the turnover is given by
$p_1$, whereas $p_2$ represents the spectral index
which should reproduce the observed X--ray synchrotron spectrum.
The normalization constant is given by $K$, giving the following general
parametric form for the injection spectrum  

\begin{equation}
Q(E)=KE^{-p_1}\left (1+(E/E_o)^{f}\right)^{\frac{p_{1}-p_{2}}{f}}.
\label{inj.eq}
\end{equation}

Particles must be accelerated beyond the light cylinder, since $\sigma$ 
must decrease significantly from a value $\sigma \gg 1$ near the
light cylinder, to a value $\sigma \sim 1$ or much less at the
pulsar wind shock (see also the discussion by Bogovalov 1999). 
A value will be derived for $\sigma$ at the pulsar
wind shock radius, $r_s$, where most of the X--ray emission is seen. 
This interpretation also holds for the magnetic axis model of RD01, since
the phased radio to X-ray emitting electrons (in the case of Vela,
Harding et al. (2002) discovered a non-thermal X-ray pulse in phase with the
radio pulse)
streaming from the magnetic polar axis
do not have enough energy per particle
to radiate synchrotron X-rays in the much weaker
magnetic field (tens of microgauss) in the plerion. However, it
remains to be shown if the linear accelerator
model of Contopoulos \& Kazanas (2002) can explain the acceleration of
electrons from mildly relativistic energies to
TeV energies, which would reduce $\sigma$ in that part of the pulsar wind.
Another problem with the model of RD01 is that the electrons streaming along the
polar axis, with magnetic footprints anchored in the open field line region,
should diverge towards the light cylinder, creating a nearly isotropic source
of electrons, rather than pencil beams at and beyond the light cylinder, as
proposed by RD01. This problem can only be overcome if it can be shown
that the acceleration of electrons to TeV energies (outside the light cylinder)
is confined to the magnetic axis of an orthogonal rotator,
sweeping across the sky, whereas the electrons
starting on field lines closer to the edge of the polar cap do not experience
the acceleration process as proposed by Contopoulos \& Kazanas (2002).

If, however, the equatorial origin for the X-ray arcs proposed by Helfand et al.
(2001) holds, which would create problems with the radio pulse interpretation
as discussed by RD01, we can still explain the bright arcs in terms of the two
co-latitudes above and below the spin equator, where the wavy neutral sheet of an
inclined rotator changes from neutral to a magnetic field with a singular
azimuthal direction. In this case the opening angle of the 
wavy neutral sheet, which is also equal to the
angular separation between the bright arcs, will depend on the angle between the 
spin axis and magnetic axis.

These apparent conflicts will have to be sorted out, which is beyond the scope
of this paper, but the results derived here can be interpreted in terms of both
models.

\subsection{The steady state electron spectrum in the compact nebula}
As stated in Section \ref{pler_intro}, we have to concentrate on the multiwavelength
spectrum of an emission volume where the transport equation can be written
in its most simplest form with minimal free parameters. This is realized in
the compact X-ray nebula where the 
convection speed at distances slightly beyond the pulsar wind shock radius,
$r_s$, is still close to $c$ (Kennel \& Coroniti 1984a, hereafter KC84.)
At least it is possible to write the convection speed in terms of $r_s$
and $\sigma$. Furthermore, it can be shown that synchrotron
losses near $r_s$ should still be unimportant
(de Jager et al. 1996a), except for energies near 30 MeV
where synchrotron losses are expected to be 
severe, resulting in a spectral cutoff (de Jager et al. 1996c).
It can also be shown that the timescale for adiabatic losses is
about three times larger than the convection timescale, if the 
radial scale size of change in the convection speed is comparable to
the width of the emission region. We can, therefore, neglect this 30\%
contribution from the adiabatic loss term (see 
de Jager \& Harding 1992 for a general expression for adiabatic losses
in terms of the gradient of the convection speed given the azimuthal
configuration of the magnetic field lines in the pulsar wind).

\begin{figure}

\plotone{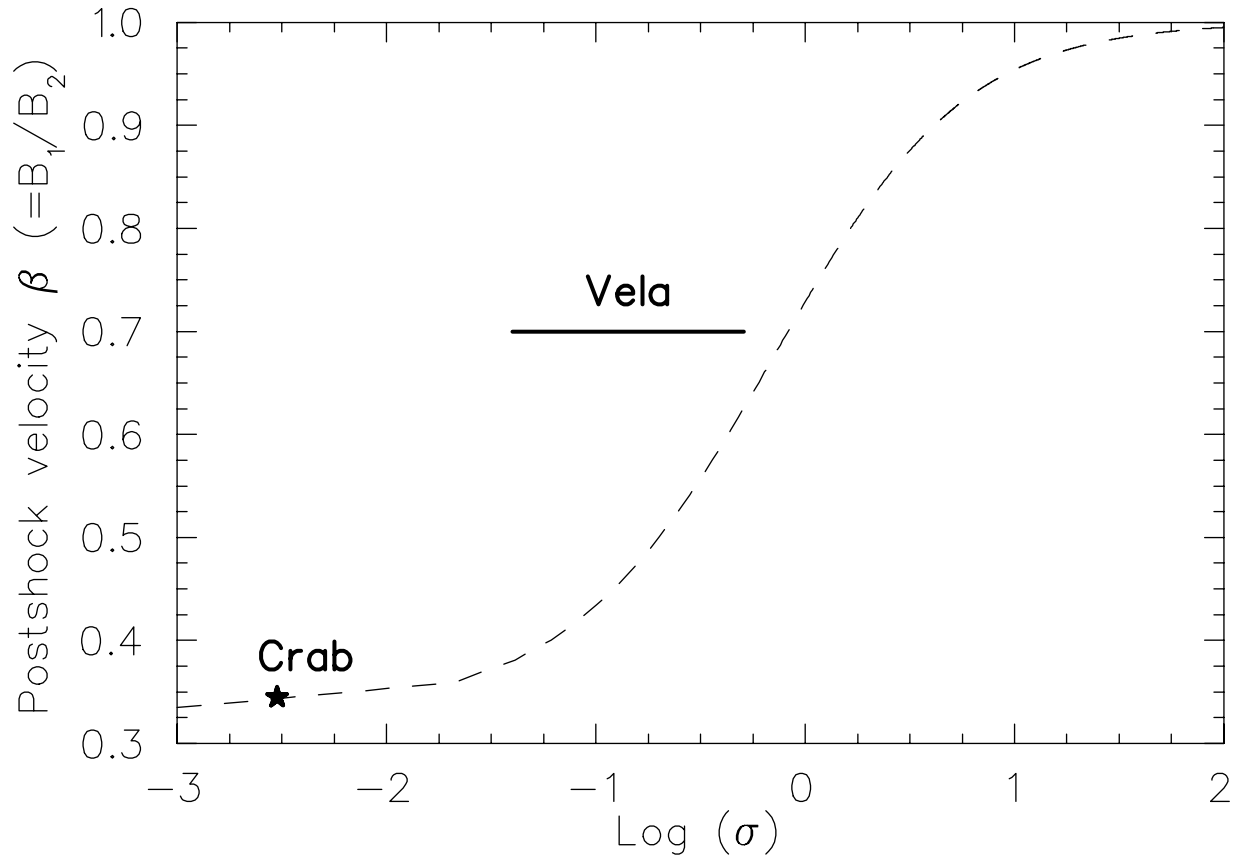}
\caption{The ratio of the upstream to downstream magnetic field strength (which
is the same as the ratio of the downstream to upstream convection speed, 
$\b=v/c$)
at the pulsar wind shock radius, $r_s$, as a function of the magnetization parameter,
$\sigma$ -- calculated from the KC84 model. The
measured convection speed for Vela is indicated by the horizontal bar ($\b$
by Helfand et al. 2001 and $\s$ in this paper). The ``star" (for Crab)
represents the theoretical value derived by KC84 and confirmed implicitly
by fitting the energy dependent shape of the X--ray torus with the e$^{\pm}$
transport equation which depends on $\b$ and $\s$ (de Jager 2001).}
\label{vc.cont}
\end{figure}

It is, therefore, sufficient to model the 
steady state electron spectrum in the region where the compact
nebula is brightest as
\[N(E)dE \sim Q(E)\tau(r_s,\sigma) dE,\] 
where $\tau(r_s,\sigma)$ is the residence time for 
electrons in the equatorial region
given by $\tau\sim \epsilon(\sigma)r_s/c$ -- 
the light travel time over a side size comparable to $r_s$. The shock radius, 
$r_s$, is assumed here since the effective width of the emission region for 
sources is $\sim r_s$.
KC84 give expressions for the convection
speed $v=c/\epsilon(\sigma)\leq c$ at $r_s$, which is 
close to $c$, but depends to some degree on $\sigma$. 
Figure \ref{vc.cont} shows the downstream
behaviour of $v/c$ at $r_s$ as a function of $\sigma$.
This ratio is also equal to the ratio of the unshocked to shocked magnetic field
strength at $r_s$. For $\sigma \ll 1$ we obtain a ratio of 1/3, consistent
with that of a strong shock, such as the Crab pulsar wind shock, 
whereas for $\sigma \gg 1$, this ratio
converges to unity, which implies an unshocked region.
The mere existence of the compact nebula implies that $\sigma$ cannot be
much larger than $\sim 1$, or else it would not have been possible to 
remove the ``cold" electrons from the co-moving pulsar wind. By shocking the
electrons, it is also possible for the electron
pitch angles to become isotropic, resulting in detectable synchrotron emission
from any viewing angle.

Even though the Kennel \& Coroniti model was employed, it is generally expected
that the convection speed should decrease beyond the pulsar wind shock. 
From Figure \ref{vc.cont}, we see that the reduction at $r_s$ is at most a factor
of 3 relative to $c$ for Crab-like plerions, for which $\sigma \ll 1$. Evidence
that the reduction in the convection speed beyond the pulsar wind shock is comparable
to the estimates derived from the Kennel \& Coroniti model is evident from the
detection of temporal changes in the
brightness of the X-ray arcs of Vela. Helfand et al. (2001) reported on the detection
of a change in the plerionic brightness
in response to energy input from a glitch which occurred
35 days ($3.0\times 10^6$ sec) before the change in plerionic brightness.
This event, which is very likely to be associated with the glitch,
required an electron  propagation speed 
of $v\sim 0.7{\rm c}$ between the pulsar and the outer arc of 
the \chandra X--ray image, assuming a distance of 250~pc from the Vela pulsar
(Helfand et al. 2001). This propagation velocity implies $\e(\s)\sim 1.4$, which
is also indicated in Figure \ref{vc.cont} and Table \ref{pul_par} (from KC84 
and the $\s$ parameter used).
 
Thus, remaining uncertainties in $v/c$ and the corresponding ratio of the
magnetic field strength between upstream and downstream should not make
a significant difference in the final results in this paper. However, if
the propagation speed was totally unknown, the results of this paper
would have depended on this unknown propagation speed through the plerion.

\subsection{The effect of geometry on the compact nebular spectrum}
The equation for the energy flux of particles in
the plane of X-ray bright emission is constrained
by the pulsar spindown power $\dot{E}$, the $\sigma$ parameter at $r_s$
and the fraction $\Omega_E/4\pi$
of the Poynting flux exploited for particle acceleration
in the pulsar wind:
\be
\int_{0}^{\infty} Q(E)EdE =\frac{\Omega_E\dot{E}}{4\pi(1+\sigma)}.
\ee
A solid angle of $\Omega_E=4\pi\sin{\theta_E}$ 
is introduced to account for the 
fact that the synchrotron emission in the compact nebula is
not isotropic, but rather confined to a torus with a half opening angle
of $\theta_E$. Helfand et al. (2001) derived a value of $\theta_E =23.3$ 
degrees
from observations of Vela, and we will assume the same value.
Thus, a fraction $\sin(\theta_E)$ of the total electromagnetic energy flux
is used to accelerate particles in the spin equator. The linear accelerator
model of Contopoulos \& Kazanas (2002, hereafter CK02) may be responsible
for this acceleration process in the spin equator. The effect of this is to 
reduce the magnetization parameter, $\sigma$, from a value $\sigma \gg 1$
near the light cylinder 
to a value $\sigma \sim 1$ or much less (as discussed previously).
A fraction $(1-\Omega_E/4\pi)$ of the spindown power
is expected to be transported to the region excluding the X-ray bright
region, which corresponds to the high latitude regions relative to the spin
equator. This region of low X-ray brightness may correspond to less efficient
acceleration, so that the wind would be magnetically dominated (i.e. high
$\sigma$ at high latitudes). 

The electron continuity equation is given by
\begin{equation}
\int_{0}^{\infty} Q(E)dE = \eta_p\frac{I_{\rm GJ}M}{e}, 
\label{flx.cns}
\end{equation}
with $\eta_p$ the fraction of all pairs escaping through the
light cylinder, which contributes to the compact nebular emission.
The expression for the Goldreich--Julian current (converted to a rate)
from the pulsar polar caps 
is $I_{\rm GJ}/e\sim\sqrt{6\dot{E}c}/e$. The pair creation multiplicity
is $M$ as defined earlier.

The continuity equation for $Q(E)$ also involves a geometrical uncertainty. 
If the particles emerge isotropically from the
pulsar, then a term $\eta_p=\Omega_p/4\pi$ (as above for the Poynting flux
conversion) has to be introduced, 
with $\Omega_p = 4\pi \sin(\theta_p)$, if only an equatorial section of the 
pulsar wind, with $\theta_p=\theta_E$, which means that the same fraction of 
the spindown
power converted to particle energy equals the fraction of pairs experiencing
this acceleration. The result is
a reduction in $\sigma$ at the spin equator, as observed in the Crab Nebula.
The possibility $\eta_p\sim 1$ in the RD01 model is implicitly
ruled out, since this would imply that all electrons
resulting from the pair cascade in the open magnetosphere have to
be confined to the two magnetic axes of the pulsar (with magnetic inclination
angle 70 degrees) as discussed previously,
resulting in two pencil beams. The reason for this is that the divergence of 
the field lines with increasing
radius above the polar cap of the neutron star does not allow such a configuration,
unless all pair cascades are concentrated on the two magnetic axes, which
is unlikely. Thus, 
considering the divergence of field lines above the polar cap, resulting
in a near-isotropic outflow of particles beyond the light cylinder, the RD01 model
can be saved if only those electrons originally escaping along the two magnetic
axes are preferentially accelerated by a linear type accelerator as suggested by
CK02. However, the CK02 model predicts acceleration in the spin equatorial
zone only, given their assumed axisymmetric pulsar magnetosphere. This model 
has to be generalized
to treat a highly inclined rotator such as Vela with inclination angle
$\sim 70$ degrees as suggested by RD01. It would then still
be a surprise if the acceleration 
process favours only those electrons coming from the two magnetic axes (as implied
by RD01), rather than those confined to the spin equator. Furthermore, even if
this is realized, the implied geometric factor $\eta_p \ll 1$, since 
only those
electrons streaming along the two poles would be selected for acceleration
(if possible). This may place a severe lower limit on the total required
pair multiplicity, $M$.  

An interesting idea also develops for outer gap geometries originally proposed
by Cheng, Ho \& Ruderman (1986): accelerated electrons from two outer gaps (resulting in a
double pulse phase separation of $< 180$ degrees), would, if accelerated by a linear
accelerator, mimic an ``offset" dipole in the plerion. RD01 found a better fit
for Vela, given such an offset dipole. However, several questions arise from
such a proposed geometry, especially the interpretation of the polarization swing for
Vela. The Vela compact X-ray nebula, nevertheless, offers an important laboratory
to study models which predict the emission of outer gap accelerated particle 
beams escaping beyond 
the light cylinder, since the pitch angles from these particle beams must
become randomized in the plerionic shocks, thus revealing otherwise hidden 
pulsed photon beams, given our viewing position.  

Given the above mentioned geometric uncertainties, we will assume a geometric factor of
$\Omega_p=6.3$, corresponding to $\theta_p = 30$ degrees for the contribution
from both poles. This is larger than the half opening angle of 23.3 degrees
derived by Helfand et al. (2001) for Vela
due to two reasons: the fainter emission
outside the two bright arcs are accounted for, and, the possibility of
a polar cap radius slightly larger than the canonical radius (due to 
inertial drag effects at the light cylinder) is also accounted for. However,
this latter uncertainty and the pair production multiplicity appears as
an irreducible normalization constant in the continuity equation.
Thus, if the effective $\theta_p$ is smaller due to whatever reason, our results
for the pair multiplicity, $M$, should be increased accordingly.
In the discussion below, it will be shown that PSR B1706-44 cannot be fitted
with a $\q_E=\theta_p=30$ degrees. In fact, a smaller value is required by the
observed spectrum. 

\subsection{The magnetic field distribution in the compact nebula}
\label{bfield}
To calculate the synchrotron emissivity we must also know the
value of the field strength at and beyond $r_s$:
the azimuthal component of the magnetic field strength in the pulsar
wind is known to scale as $B(r)\propto 1/r$, up to the radius
$r_s$ where the ram pressure from the pulsar wind is balanced
by the external pressure (static or dynamic gas pressure, whichever
term dominates). The amount of field
compression at $r_s$ will depend on the 
strength of the shock. For strong Crab--like relativistic shocks 
($\sigma \ll 1$) the compression factor is $\kappa(\s) \sim 3$, whereas
smaller factors are expected for weaker shocks (larger $\sigma$). 
From KC84, we find that the ratio of upstream to downstream 
field strengths at $r_s$ is given by 
$1/\kappa = v/c=1/\e(\s)$, as shown in Figure \ref{vc.cont}. Thus, 
a shock (corresponding to $\sigma <1$) is also associated with 
a jump in the field strength across $r_s$.  The postshock field strength
at $r_s$ for a given $\sigma$ is given by KC84 as 

\begin{equation}
B(r_s)\sim \kappa (\sigma)\left[\frac{\dot{E}}{cr_s^{2}}\frac{\sigma}{(1+\sigma)}
\right]^{1/2}.
\end{equation}

\begin{figure}

\plotone{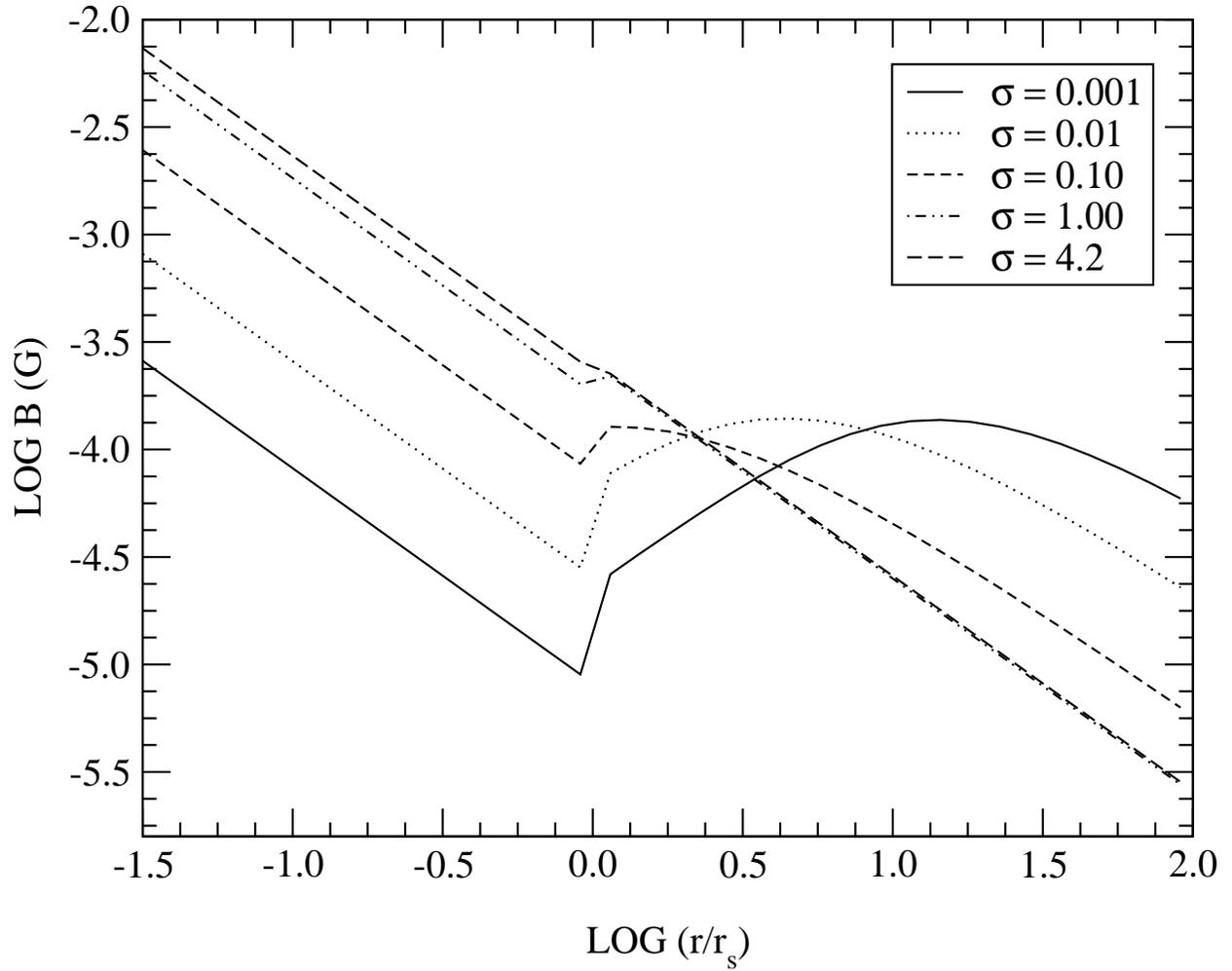}
\caption{The magnitude of the azimuthal component of the
magnetic field strength as a function of radius in the
pre--shock ($r<r_s$) and post-shock ($r>r_s$) regions of the pulsar wind
for $\sigma$ as indicated (as derived from KC84).}
\label{br.cont}
\end{figure}

This expression shows the effect of particle energy flux domination for
low $\sigma$ shocks, since $B(r_s)$ approaches zero asymptotically as 
derived by KC84. The no-shock field dominated
case ($\sigma \gg 1$) gives the well known limiting expression 
$B(r_s)=(\dot{E}/c)^{1/2}/r_s$. Graphs for $B(r)$ in the low $\sigma$-limit
are shown in Figure \ref{br.cont} for the upstream ($r<r_s$) and downstream
($r>r_s$) regions, employing the spindown parameter of PSR B1706-44, and
assuming an inner pulsar wind shock radius of $r_s=1$ arcsec -- close to
the value inferred by Dodson \& Golap (2002). This graph was plotted to 
a radius corresponding to the mean radius of $1.'5$ 
measured for the compact radio nebula by Giacani et al. (2001). A
termination shock is expected at that position, resulting from the static
gas pressure.

For both Vela and PSR B1706-44 we find that the width of each X-ray compact
nebula is comparable in size to $r_s$ (ignoring the effect of trails
due to proper motion). Self consistency with the expected postshock field
distribution (assuming the KC84 low-$\sigma$ solutions for $B(r)$)
implies that the width of the emission volume in the bright arcs
of Vela and the size of the X-ray compact nebula of PSR B1706-44 should be
consistent with the gradient in the magnetic field distribution in the 
postshock region.
Using a numeric approach, de Jager (2001) proved this consistency between 
the energy dependent
shape of the X-ray torus and the KC84 field distribution for $\sigma=0.003$ --
the preferred magnetization parameter for the Crab pulsar wind shock derived
by KC84. We assume that this general formalism also holds true for the other plerions.

In the KC84 formalism, we find that $\sigma\sim (r_s/r_N)^2$, where
$r_N$ is the radius of the outer edge (termination shock) of the compact nebula. For the
Crab Nebula $r_N\sim 200$ arcsec and $r_s\sim 10''$, which correctly 
predicts $\sigma=0.003$,
even though the radius of the X-ray torus is significantly smaller than
$r_N$ for the Crab. This can be understood in terms of synchrotron burnoff
effects (KC84, de Jager 2001). In the case of Vela and PSR B1706-44, the radial
scale size of the X--ray nebulae is more than an order of magnitude smaller than the
corresponding scale size of the Crab X--ray nebula, in which case the electron
residence time and hence the synchrotron burnoff timescales for these two 
sources are also much less compared to that of the Crab X--ray emitting 
electrons. Thus, the large disparity in radio and X--ray sizes are not the
effect of synchrotron losses, but has to do with differences in the field
strengths as discussed below. A factor ten drop in $B$ would correspond 
to a similar drop in the synchrotron frequency, but a factor 100 drop in 
intensity, thus redshifting
the 0.5 keV - 8 keV \chandra band into the EUV/soft X--ray band, 
rendering the X-rays
faint or undetectable in the radio nebula. This reduction in $B$ is
indeed anticipated by the KC84 model. Figure \ref{br.cont} shows 
indeed a drop by
a factor of 10 between $B(r_s)$ and $r=90$ arcsec ($\log(r/r_s)=2$), corresponding to
the size of the radio nebula for PSR B1706-44.

Dodson \& Golap (2002) already had to invoke a relatively high pressure 
to account for $r_s=1$ arcsec, which should be kept in mind for the discussion below.
The observed RMS scale size of the compact nebula allows us now
to constrain the value of $\sigma$, given Figure \ref{br.cont}. Given an RMS
scale size comparable to $r_s$ itself, the scale size over which the magnetic field
decreases must also be comparable to $r_s$, which corresponds to values of 
$\sigma > 0.05$. If, due to some reason, $\sigma$ was smaller, say
$\sigma \sim 0.01$, as shown in Figure \ref{br.cont}, we find that $B(r)$ must increase
at $r>r_s$, until equipartition is met, after which $B(r)$ starts to drop again.
Such a Crab-like field distribution would imply an emission region for which
the width is at least 10 times the size of $r_s$ (defined to be the region
for which $B(r)\geq B_s$, with $B_s$ the postshock field strength at $r_s$),
which is ruled out by \chandra observations of this source. 
If, due to whatever reason, we still require that 
$\sigma$ must be as low as 0.01 (the example shown in Figure \ref{br.cont}),
or even lower, the observed width of the emission
region would force us to declare a value of $r_s = 0.1$ arcsec (unresolved by
{\it Chandra}). The magnitude of the confining gas pressure 
and/or pulsar proper motion requirements would then be nearly impossible to
meet. Thus, keeping $r_s\sim 1$ arcsec for PSR B1706-44, we have to
assume that $\sigma$ should be at least more than 0.05 (to avoid a too wide
emission scale radius, not supported by observations), but not much
larger than unity (to maintain the properties of a shock for the formation
of the compact nebular emission). 

Similar arguments can be applied to the Vela bright arcs - their structure also
implies a similar range for $\sigma$, but we will see if this constraint
can be independently derived from the observed compact nebular
multiwavelength spectrum as discussed below.

All the aforementioned discussions were required to motivate our use of $B_s$
in the calculations of the synchrotron emissivity. In fact, we use a value for
$B$ which is averaged over the observable size of the compact nebular emission.
This average was found to be $\sim B_s$ and the latter will be used in the
calculations presented below.
Our treatment for the field strength is thus relatively accurate, given the
arguments above.

\section{Constraints on the pair multiplicity and wind magnetization
parameter}
\label{pler_pair}
The expected compact nebular synchrotron spectra were calculated
from the steady state particle spectrum, $N(E)$, in the compact nebula
and the spatially averaged field strength $<B(r\sim r_s)>\sim B_s$ 
for Vela and PSR B1706--44, following the procedure in the preceding
sections. Comparison
with the measured spectra (including limits where appropriate, see Figure
\ref{vela.1706})
resulted in significant constraints for $\sigma$ and $M$ as discussed below. 

To arrive at a steady particle spectrum, $N(E)$, we have to employ the
residence time, $\tau(r_s,\sigma)$, for electrons (discussed previously)
in the emission zone for which the scale size is $r_s$.
Thus, given all arguments discussed above, 
$N(E)\sim Q(E)\tau(r_s,\sigma)$
as discussed previously. The electron spectral index $p_2$ 
above the turnover energy ($E_0$) and its
uncertainty was derived from the power law index of the 
observed compact nebular X--ray spectrum. The lower energy index
$p_1$ (below the turnover energy) is typically constrained
by radio and optical observations (or their upper limits), as well
as by the condition to have both the energy and continuity
equations satisfied, resulting in acceptable and forbidden parameter spaces.
Furthermore, all synchrotron observations and their upper limits
had to be fitted. All free parameters: $p_1$, the error
on $p_2$ (derived from the uncertainty on the X--ray
spectral index), $E_0$, $f$, $d$ (distance to the pulsar)
and $E_{\rm max}$, were varied within their confidence limits, given each
choice for $M$ and $\sigma$. In the case of statistical errors on measured
parameters, a search was made within their 68\% confidence interval.
A minimum Chi-squared ($\c^2$) statistic was calculated 
for each choice of $M$ and $\sigma$, and up to the 1-sigma 
confidence contours (derived from the $\chi^2$ statistic)
were constructed as shown in Figure \ref{sig_mult}. The relevant observations, 
with derived results, are given in the following two subsections on Vela and 
PSR B1706-44.

\subsection{Parameter constraints for Vela}
\label{pler_pair_vla}
Whereas the spindown power, $\dot{E}$, is known for each pulsar, the
external gas pressure, $P_{\rm gas}$, is known only for Vela, 
from which the observed value of $r_s\sim 33$ arcsec was 
derived by Helfand et al. (2001). This value, used in the calculation
of the expected compact nebular synchrotron spectrum,
is relatively close
to the inner radii of the two bright arcs derived by RD01. 
The field strength of the compact nebula $B(r_s)$ is, therefore, also known,
given a value for $\sigma$. 
The multiwavelength spectrum employed for Vela is shown in Figure \ref{vela.1706},
with details in the caption as well as the discussion given below.

\begin{figure}
\plotone{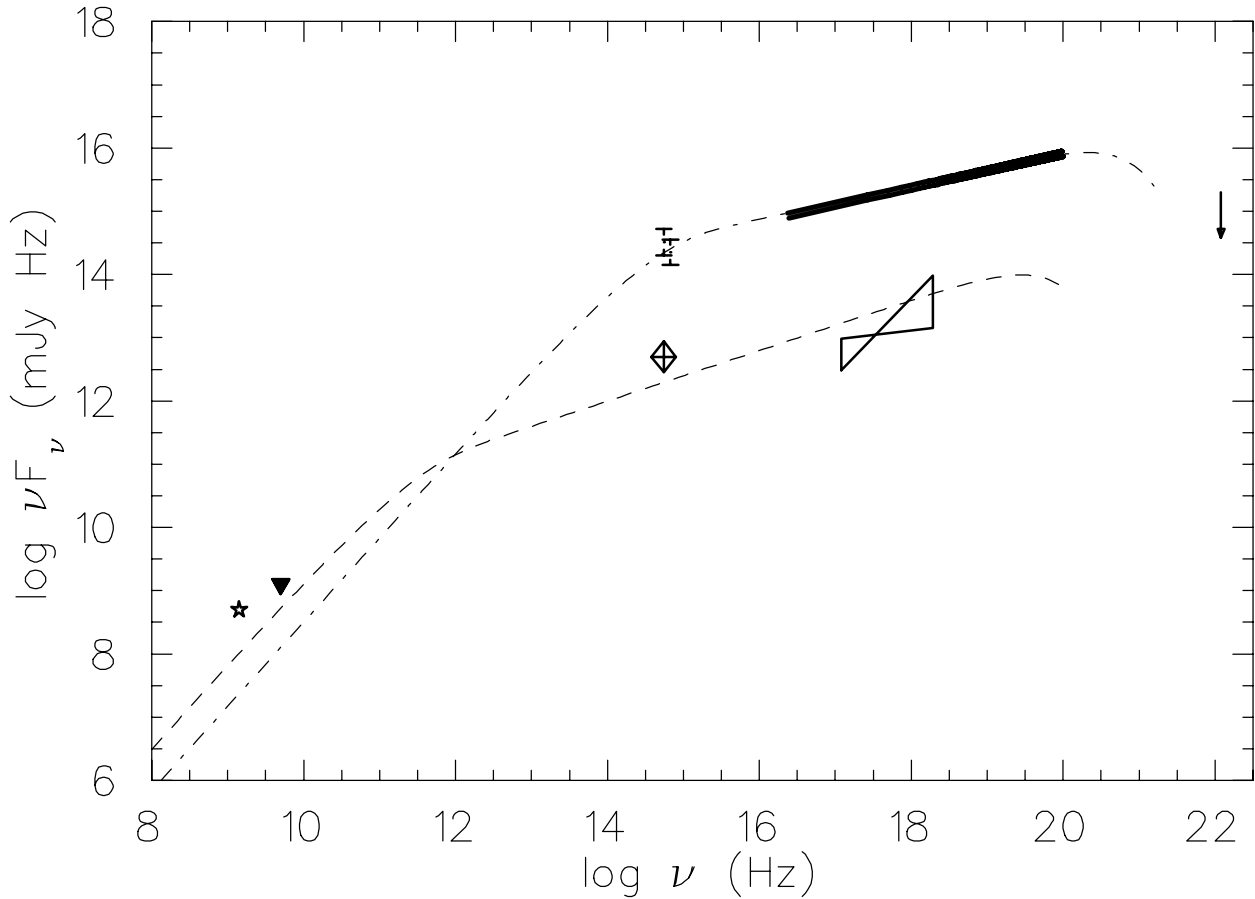}
\caption{The Vela and PSR B1706--44 compact nebular synchrotron 
spectra from the observations and the model calculations. Vela: The closed 
triangle shows the upper limit to the compact radio nebula, 
whereas the two vertical bars represent the V and B band (upper limits) fluxes 
measured by \"Ogelman et al. (1989). 
The thick double-line corresponds to the best fit X-ray compact nebular
spectrum measured up to 0.4 MeV (de Jager et al.
1996a), and the arrow shows the upper limit to EGRET detection. The 
dot--dashed line corresponds to the best fit synchrotron
spectrum for an equatorial disk with parameters given in the text. 
PSR B1706--44: The star represents the upper limit to the off--pulse 
sensitivity inside $91''\times10''$ (Gaensler et al. 2000); the diamond 
indicates the upper limit to the optical flux (Lundqvist et al. 1999), and the
1-sigma X--ray error box is the \chandra detection between 0.5 and 8.0 keV 
(Gotthelf
et al. 2002). The dashed
line represents the compact nebular synchrotron spectrum for the torus (from 
the model with constraints).} 
\label{vela.1706}
\end{figure}

Even though Lewis et al. (2001) measured a 6 cm radio flux
of 0.26 Jy in the two radio lobes perpendicular to the 
spin axis/proper motion vector of the pulsar, almost all this
emission lies outside the X--ray compact nebula. The analysis
in this Section, however, isolates the convective 
flow in the compact nebula, so
that the radio flux of $\sim 0.25$~mJy at 6 cm  
(given by Bietenholz, Frail \& Hankins 1991),
which corresponds to the edge of the compact X--ray nebula,
is considered as a 3-sigma upper limit for emission from inside
the compact nebula. The reason why this flux is considered
as an upper limit is because some of the diffuse and/or
structured radio emission seen superimposed on the compact nebula 
may be the result of foreground or background emission from the
radio lobes of Lewis et al. (2001). 

Optical B- and V-band fluxes derived by \"Ogelman et al. (1989) are used as 
upper limits to constrain the turnover energy $E_o$ and these are consistent with
the extrapolation from the compact nebular spectrum derived by de Jager et al.
(1996a) between soft X-rays and low energy $\gamma$-rays. 

The X--ray spectrum used for the
fit procedure was taken from de Jager et al. (1996a) and Strickman, de Jager
\& Harding (1996) with a
relatively small error on the power law spectral index (photon
index of 1.73) given the broadband nature of the 
fit, extending between soft X--rays and low energy $\gamma$--rays.
EGRET observations (de Jager et al. 1996b; de Jager 1995) 
appear to constrain this hard component, requiring a 
high energy turnover or cutoff, which is accounted for
by the maximum electron energy, $E_{\rm max}$, which is treated
as a free parameter. The spectrum must, however,
turn over around or below the $R$-band to be consistent with the
radio upper limits.

The model synchrotron flux was calculated at those frequencies for
which constraining data (with their errors) were available.
The quality of the fit was quantified by calculating a $\chi^2$
statistic, as discussed previously, 
which measures the difference between the predicted
and observed fluxes at various energies, but weighed properly 
with the 1-sigma error on each observed flux.  
In the case of a typical 3-sigma upper limit (UL), 
the contribution to the 
$\chi^2$ statistic was calculated by adding the term 
$(3F_{\rm pred}/{\rm UL})^2$ ($F_{\rm pred}$ is the predicted flux), 
which gives a zero contribution if the predicted flux is well below
the observed flux, but 9 if the predicted flux is equal to
a 3-sigma upper limit. 

The 1-sigma confidence contours for $M$ vs
$\sigma$ are plotted in Figure \ref{sig_mult} (while keeping the other
parameters free) using the prescription
of Lampton, Margon \& Bowyer (1976), which involves plotting the contours
corresponding to $\chi^2({\rm min})+S$ with $S=2.3$ corresponding
to a 1-sigma and $S=4.6$ for a 1.6-sigma confidence contour.

\begin{table}
\begin{center}
\caption{Acceptable sets of model parameters of the synchrotron nebulae of 
PSR B1706--44 and Vela for the purpose of spectral fitting.The given set is 
for the toroidal emissions from the two systems.}
\label{pul_par}
\begin{tabular}{l l l l}
 & & &\\
\hline
Parameter&Description &  B1706--44& ~Vela\\
\hline
P  &pulsar period (ms)                          & 102.4 &  89.3   \\
$\dot{E}$ &spindown power ($10^{36}$ erg/s)&  3.4  & 6.9 \\
$\tau_{age}$ &characteristic age ($10^3$ yrs)     & 17.5  &  11.5\\
d &distance to the pulsar/plerion (pc)  & 3200  &295  \\
$\q_E=\q_p$ &half opening angle (degrees)    &15     &30\\
$P_{gas}$&thermal gas pressure ($10^{-10}$ erg~cm$^{-3}$)&33$^a$    & 8.5\\
$p_1$&electron spectral index ($E<E_0$) & -1.0     & -1.0\\
$p_2$&electron spectral index (X-rays)&2.2    &2.46\\
f&fastness of the turnover (at $E_0$) &10     &10 \\
$r_s$ &shock radius (arcsec) & 1.1    & 33\\
$r_s$ &shock radius ($10^{17}$ cm)  &0.53	 &1.47\\   
$B(r_s)$&magnetic field strength at $r_s$ $(10^{-4}\; {\rm G})$ &1.4    &0.72\\
$\sigma$&magnetization parameter at $r_s$     &0.1    &0.1\\
$\epsilon(\sigma)$&convection speed ($v=c/\epsilon(\sigma)$)& 2.3   & 2.3\\
$\kappa(\sigma)$&$B(r\ge r_s)$/B($r < r_s$)&2.3      &2.3\\
$M$&pair creation multiplicity & $10^4$    & 501\\
$E_{0}$ &minimum electron energy (ergs)&0.03   &1.3\\
$E_{max}$ &maximum cutoff energy (ergs)  &150    &650\\

\hline
\end{tabular}
\end{center}
\noindent $^a$ Such a static pressure would reproduce $r_s=1''$, although
\noindent the pressure of a $5''$ tail requires $nv_{150}^2\sim 10$ as derived
\noindent by Dodson \& Golap (2002). 
\end{table}

The Vela contours close and it 
is clear that $\sigma$ lies between 0.05 and 0.5 for Vela, consistent
with the constraint $0.05<\sigma <1$ derived from an independent argument
in Section \ref{bfield}. This is slightly lower than the value
near unity considered by de Jager et al. (1996a) 
based on phenomenological arguments. 
This is, however, the first direct measurement of $\sigma$
for Vela, based on robust model assumptions. The multiplicity,
$M$, above the polar cap is between 300 and 1000. 
The best-fit parameters for Vela (and PSR B1706--44) are given in 
Table \ref{pul_par}.

\subsection{Parameter constraints for PSR B1706--44}
\label{pler_pair_b17}
A pulsar wind shock radius was fixed at $r_s=1$ arcsec,
which implies (Dodson \& Golap 2002) $nv_{150}^2\sim 10$ 
(with $n$ the gas density and $v_{150}$ the pulsar proper motion in
units of 150 km/s) given a distance
of 3 kpc. The dynamic pressure (due to proper motion), rather than the static
pressure was used, since the image of PSR B1706-44 shows a cometary tail, indicating
that the pressure due to proper motion exceeds the static gas pressure.
The ROSAT data implies $n\sim 0.09$ cm$^{-3}$ (Dodson 1997) and if one attempts
to reduce $r_s$ even more, the constraint on the proper motion would become
severe.

The observed spectrum of the compact nebula in
PSR B1706--44 is also shown in Figure
\ref{vela.1706}. The radio observations show data point at 1425 MHz 
obtained from a VLA beam size limit of $91''\times10''$ (Gaensler et al. 2000),
which is still large compared to the \chandra size of the compact nebula. 
The off--pulse VLA flux limit of 0.36 mJy (per beam) at 1.4 GHz (Gaensler et
al. 2000) is used in Fig. \ref{vela.1706} as the radio upper limit to the 
X--ray compact nebula, because the size of the \chandra X--ray
nebula is small compared to the VLA beam size used for this object.  
  
The $V$-band upper limit obtained from the discussion in Section \ref{pler_1706}
is also shown in Figure \ref{vela.1706}. This limit does not constrain
the extrapolated X-ray compact nebular spectrum. A softer X-ray
spectrum (such as Vela) would have been well constrained by this limit.
The result of this is that the turnover energy $E_0$, which
is an important parameter to consider when attempting to constrain
$M$ and $\sigma$, is not well constrained. This probably explains why the
confidence contours for $M$ vs $\sigma$ do not close for this source,
as discussed below.

\begin{figure}

\plotone{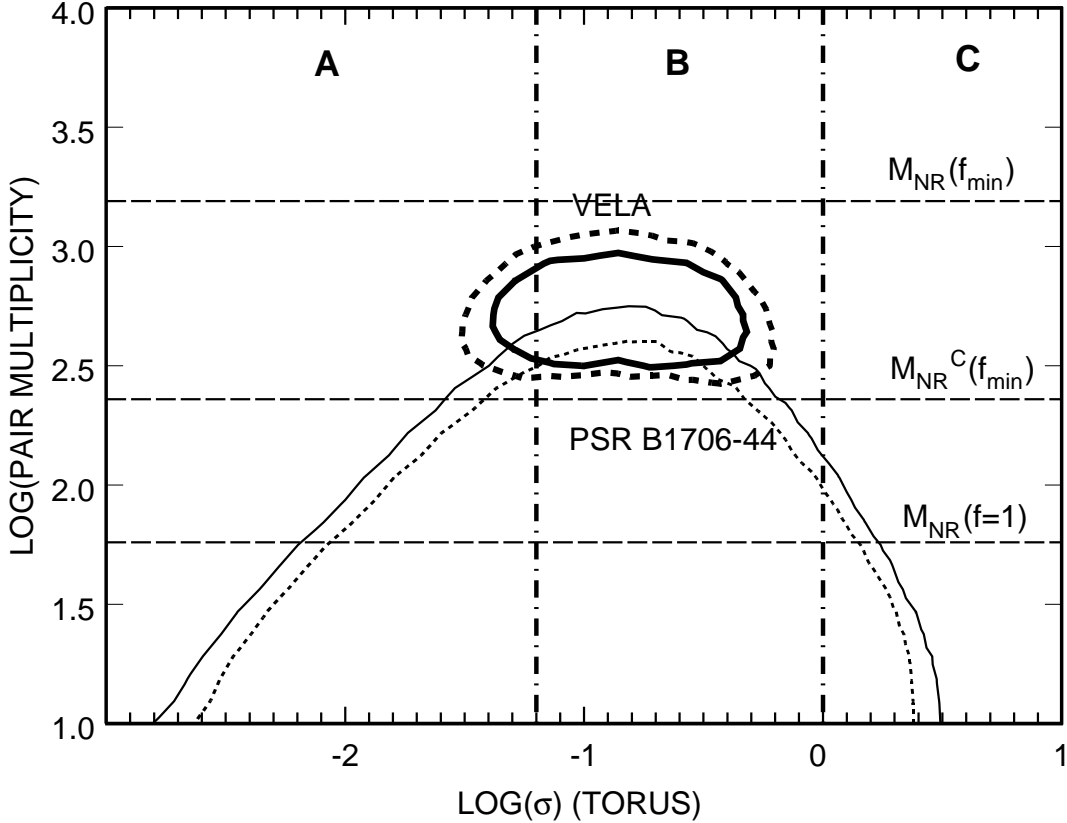}

\caption{The multiplicity-sigma relationship of the Vela and PSR B1706-44
pulsar systems. Dotted contours are at the 1-sigma level, whereas solid 
contours are at the lower sigma level.
The two vertical lines divide the $\sigma$ parameter space in three sections:
``A": strong Crab-like shocks with extended tori as seen from Crab;
``B": weaker Vela-like shocks
showing narrower emission regions with the width of the emission region
comparable in size to the pulsar wind shock radius; 
``C": The no-shock region implying no compact nebular emission.  
The predicted pair multiplicity from the Hibschman \& Arons (2001) 
model for neutron star surface temperatures of 
$T=1.6\times 10^6$ K are shown as horizontal dashed lines for the following
conditions (NRICS apply to all cases): 
$M_{\rm NR}(f_{\rm min})$ -- NRICS from a stellar surface with uniform
temperature and maximal multipolar components; 
$M_{\rm NR}(f=1)$ -- the same as above, but for a pure dipolar field;
$M_{\rm NR}^C(f_{\rm min})$ -- NRICS on soft X-ray photons from 
canonical size polar
caps at a temperature of 1.6 MK, but with maximal multi-polar structure.}
\label{sig_mult}
\end{figure}

The 90\% confidence interval on the \chandra spectral index was
employed for the uncertainty in $p_2$, and it will become clear that 
the relatively hard X--ray spectra (supported by {\it Chandra}), cannot be 
explained with this model.
The fact that we could not force a low energy turnover for the
hard compact nebular spectrum as measured by \chandra (see Figure \ref{vela.1706}),
results in failure to close the confidence contours as shown in Figure \ref{sig_mult}.
We have, therefore, to rely on the arguments given in Section \ref{bfield}
stating that $0.05 < \sigma \le 1$ based on the RMS size of the the X--ray
nebula. This constraint is indicated as
two vertical lines in Figure \ref{sig_mult}, which also implies that $M>100$
given the location of the confidence contours.

The EGRET upper limit for the unpulsed $\gamma$--ray component
from PSR B1706--44 is also not constraining relative to an extrapolation
of the X--ray spectrum, which means that $E_{\rm max}$ is unconstrained,
except by the maximum allowable polar
cap potential as discussed by de Jager at al. (1996a) for Vela.

A set of acceptable model parameters inside the 1-sigma confidence contours
was selected to plot the model spectrum as shown in Figure \ref{vela.1706}. The
corresponding parameters are shown in Table \ref{pul_par}. Note that the faint
X-ray emission in the compact nebula forces the solutions towards softer spectra
allowed by the confidence interval on the X-ray spectral index, $\alpha$, (indicated in the
Table as the particle index, $p_2=2\alpha+1$). 
Harder spectra can be explained only
if $\theta_E$ is made smaller, or, if $d$ is increased beyond 3 kpc. Relaxing
the former constraint may be more reasonable, but at the expense of future
attempts to explain the unpulsed TeV emission! A low energy turnover of the
spectrum is indeed seen, but this constraint arose from the constraint that
both the energy and continuity equations (assuming $\W_E = \W_p$) had to be 
satisfied, rather than from the radio constraint shown.

\section{Discussion}
\label{pler_disc}
The multiwavelength data for Vela
are consistent with an extension of the compact nebular spectrum from low energy
$\gamma$-rays, at least down to the V-band, providing an important constraint
for the nebular turnover energy $E_0$. This turnover must therefore be between
radio and optical energies - if the optical emission from Vela nebula claimed
by \"Ogelman et al. (1989) are indeed of synchrotron origin.

The unabsorbed X-ray luminosities of Vela and PSR B1706-44 are equal
in the \chandra 0.5 to 8 keV band if we assume a distance of 260 pc
for Vela and a distance of 3 kpc for PSR B1706-44. The scaled pulsar wind shock
radii are also the same if we assume the same confining pressure for both
sources. Scaling the 
radii of the two Vela X-ray arcs (22 arcsec and 29 arcsec) derived by
RD01 to the distance of PSR B1706-44 (accounting for the differences
in spindown power in the proper way), 
results in a predicted radius between 1.3 arcsec and 1.8 arcsec.
This is remarkably close to our assumed value of 1 arcsec, but at the expense
of assuming the same confining pressure. In the case of PSR B1706-44
we need a relatively large combination of gas density and proper motion
speed ($nv_{150}^2\sim 10$) to maintain this compactness. A discussion of
the merit of such a high value for the product of $n$ and $v_{150}$
is beyond the scope of this paper.

The X-ray spectrum of PSR B1706-44 is harder compared to that of
Vela, which made it difficult to find optical data which would
constrain the turnover energy,
$E_0$. The result was that we could not constrain the combination of
$M$ and $\sigma$ properly, resulting in open contours for PSR B1706-44. 
However, the low energy turnover for Vela, even though not well-defined,
results in a relatively well-defined solution for $M$ and $\sigma$ - despite
other parameter uncertainties. Hopefully more sensitive upper limits for PSR B1706-44
can be derived from narrow-beam radio off-pulse measurement and more 
sensitive optical observations to force a low energy
turnover in the extrapolated hard X-ray spectrum.

We were able to impose constraints on $\sigma$ for both sources. The value of
$\sigma$ in the X-ray compact nebulae of both sources is expected to lie 
between 0.05 and 1, based on the observed radial gradient of the X-ray 
compact nebular emission for both sources. Furthermore, the same interval 
(this time from 0.05 to 0.5) was independently
established for Vela after exploiting the
multiwavelength spectrum in a robust injection model. 
It is thus encouraging to see that two independent
approaches give the same interval for $\s$. We also see direct evidence 
that $\sigma$
has been reduced from a value $\gg 1$ (expected near the light cylinder)
to a value $<1$ at the pulsar wind shock radius, as expected from theoretical
considerations.

A measurement of the turnover energy, $E_0$, is also important from a physics 
viewpoint.
If pairs from the magnetospheric cascade process are transported to the pulsar 
wind shock without additional acceleration, $E_0$ should reflect the final
electron energy emerging from the pulsar magnetosphere. If, however, $E_0$ 
exceeds the average emergent energy of the electrons, 
additional acceleration of the total pair population between the light 
cylinder and the pulsar wind shock should be considered. The model of CK02 
considered this,
but possibly only in the spin equator, rather than the latitudes
corresponding to the two magnetic poles. It remains to be shown if
the linear accelerator (CK02) can preferentially accelerate the 
electrons from the polar axes
of an inclined rotator, thus creating the odd arcs in the sky observed from 
Vela, as required by RD01.

The measured neutron star temperatures for Vela and PSR B1706-44 are both
around 1.6 MK (Pavlov et al. 2001; Gotthelf et al. 2002).
This temperature corresponds to the pulsed components for both sources
and the emission radii involved are much larger than the canonical polar cap
radii. Given
the similarity of spin parameters, field strength and temperatures between Vela
and PSR B1706-44, we find that curvature losses give the minimum Lorentz factor
in the Hibschman \& Arons (2002) model. However, the pair multiplicity, $M$,
predicted for Vela exceeds $10^4$, given curvature losses as the dominant 
process, whereas the confidence contour of $M$ vs $\sigma$ for Vela implies 
values for $M$ ranging
between 300 and 1000, which are well below that predicted by curvature 
radiation. This indicates that
non-resonant inverse Compton scattering may be more important for Vela as indicated
by the horizontal lines in Figure \ref{sig_mult}. From Hibschman \& Arons (2002)
we find for a star with Vela-like parameters and a uniform surface temperature
of 1.6 MK, pair multiplicities due to non-resonant inverse Compton scattering (NRICS)
between 50 (pure dipolar structure) and 1600 (multipolar
structure above the polar cap), which are quite consistent with our measurements.
If only the canonical size of the polar cap were at this temperature, then we
find that not even NRICS in a multipolar field structure
would have been capable to explain the observations (indicated by the maximal line 
$``M_{\rm NR}^C(f_{\rm min})"$ in Figure \ref{sig_mult} for a hot cap with
canonical size). The radius corresponding to the detected pulsed emission is, however, a 
significant fraction of the stellar radius (as discussed above),
indicating that the calculations
corresponding to a uniform surface temperature is more appropriate.
There is another good reason for assuming NRICS rather than curvature losses
for Vela: Hibschman \& Arons (2002) neglected the presence of the optical pulsed 
emission seen from
Vela and the cross section for NRICS of beam electrons
on these optical photons is larger compared to
the cross section on soft X-rays from the hot surface. This may result in NRICS
on optical photons as the dominant process (relative to curvature losses.)
The predicted pair production multiplicities will then also change.

It also remains, then, to be shown if the final beam Lorentz factor along the 
polar axes (given NRICS
on the pulsed optical photons in Vela) is large enough to
reach and maintain at least 50 TeV up to the light cylinder, since this energy
is required by RD01 to explain the bright arcs without any acceleration in the plerion.

Future studies will see the combination of pulsar magnetospheric studies,
radio pulse properties, plerionic geometry and particle acceleration outside
the light cylinder much more closely related -- we have reached the level
where the one field of study can no longer survive without the other.

\begin{acknowledgements}
RRS would like to thank the National Research Foundation and the Unit 
for Space Physics at 
the Potchefstroom 
University for their financial and 
technical support during the course of his PhD. 
Useful discussions with J.P. Finley, E. V. Gotthelf, D. Kazanas, J.A. 
Hibschman, J. Arons, J. S. Gallagher and H. \"Ogelman are acknowledged. 
Helpful comments from the referee are also acknowledged.

\end{acknowledgements}

\def \refer #1 {\bibitem{#1}}

\end{document}